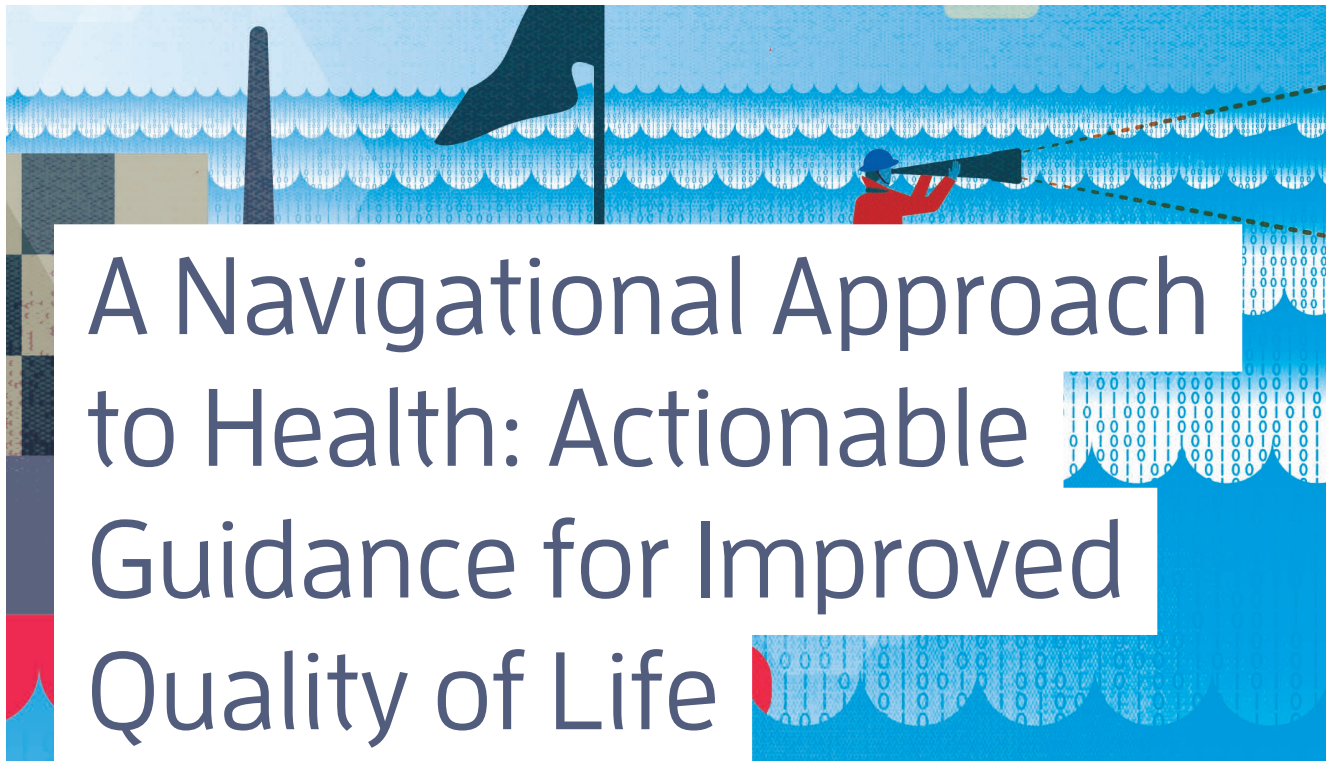

# A Navigational Approach to Health: Actionable Guidance for Improved Quality of Life

**Nitish Nag and Ramesh Jain,** University of California, Irvine

*Health and well-being are shaped by how lifestyle and the environment interact with biological machines. A navigational paradigm can help users reach a specific health goal by using constantly captured measurements to estimate how their health is continuously changing and provide actionable guidance.*

Experiences are valuable for determining a person's quality of life. If we define quality of life as a function of positive experiences and time, then we want to optimize those experiences and extend their duration. An illness can have a negative effect, despite counter mechanisms such as medications or surgery. The challenge for the next century is to ensure long lives and maintained health, thus enabling people to enjoy high-quality experiences.

In the last century, progress in health care resulted in a precipitous decrease in infectious diseases. Antibiotics reflected one side of the coin, by treating patients reactively, once an infection had already become a major health concern. On the flip side, advances in preventing infections through population-level data offered insight into ways to improve public hygiene and contain infectious agents, while vaccines biologically prevented infections. As humans continue to lead longer lives, chronic diseases have emerged as the main health challenge. Modern medicine relies mostly on reactive health systems that take care of individuals only after they become sick, using drugs, surgeries, and other "fix it when it breaks" episodic intervention methods. Prescribing antibiotics is the hallowed strategy medical experts have taken toward treating infectious disease. To tackle chronic disease and significantly improve quality of life, it is necessary to move beyond the episodic health model.

The World Health Organization defines health as "a state of complete physical, mental and social well-being and not merely the absence of disease or infirmity."[1] Health is a dynamic state that is constantly changing based on biology, the environment, and lifestyle. The state space of health is complex and multidimensional but, for the sake of simplicity, in Figure 1 we describe a one-dimensional continuum ranging from optimal wellness to death. Everyone is somewhere on this continuum at every moment, but quality of life increases if there is a leftward shift.

Technology that senses a change in a user's position on the health continuum can drastically improve human health. In prehistoric times, only symptoms (perceptual indications) guided people's actions. Ancient civilizations started to document medical conditions with basic human sensory abilities, such as how people looked, their pulse rhythmicity, coughing sounds, or the taste of urine (e.g., diabetes). Symptoms are usually latent signals of biological dysfunction. Today, advanced methods detect changes in health, but only in a professional health-care setting after a patient visits due to symptoms. Reducing the lag time in recognizing health state changes requires transitioning from the current approach to a new high-resolution multimodal continuous-sensing paradigm.

Daily decisions that affect health can result from hedonistic tendencies. The difference between finances and health is that quantitative health status is essentially invisible to us, especially when we are healthy. In financial matters, people can quantitatively see how much they are saving or spending at any given moment. They can choose to save for the future or spend. If people become aware of their biology, they could see how daily life is affecting their health and make informed decisions accordingly. Health decisions are too important to be episodic; they must be an intrinsic part of daily life. This is the motivation behind proposing the navigation perspective.

## DEFINING PERSONAL HEALTH NAVIGATION

Navigation is different from recommendation systems and automatic circuits. This goal-based guidance system perpetually estimates the current state, computes the best route through intermediate states, and guides actions that lead to health goals. For true success, this system requires robust continuous sensing, accurate estimation of the current state, control systems, actuation mechanisms, understanding of the outside world and context, long-term planning, and goal decomposition at various levels of detail. Navigation must also manage stochastic, chaotic, and noisy environments effectively. This goal-driven, closed-loop sense-compute-then-act cycle combines the pioneering elements from cybernetic systems (well described in electromechanical systems and biology by Weiner et al.[2]) and the general problem solver by Newell et al.[3]

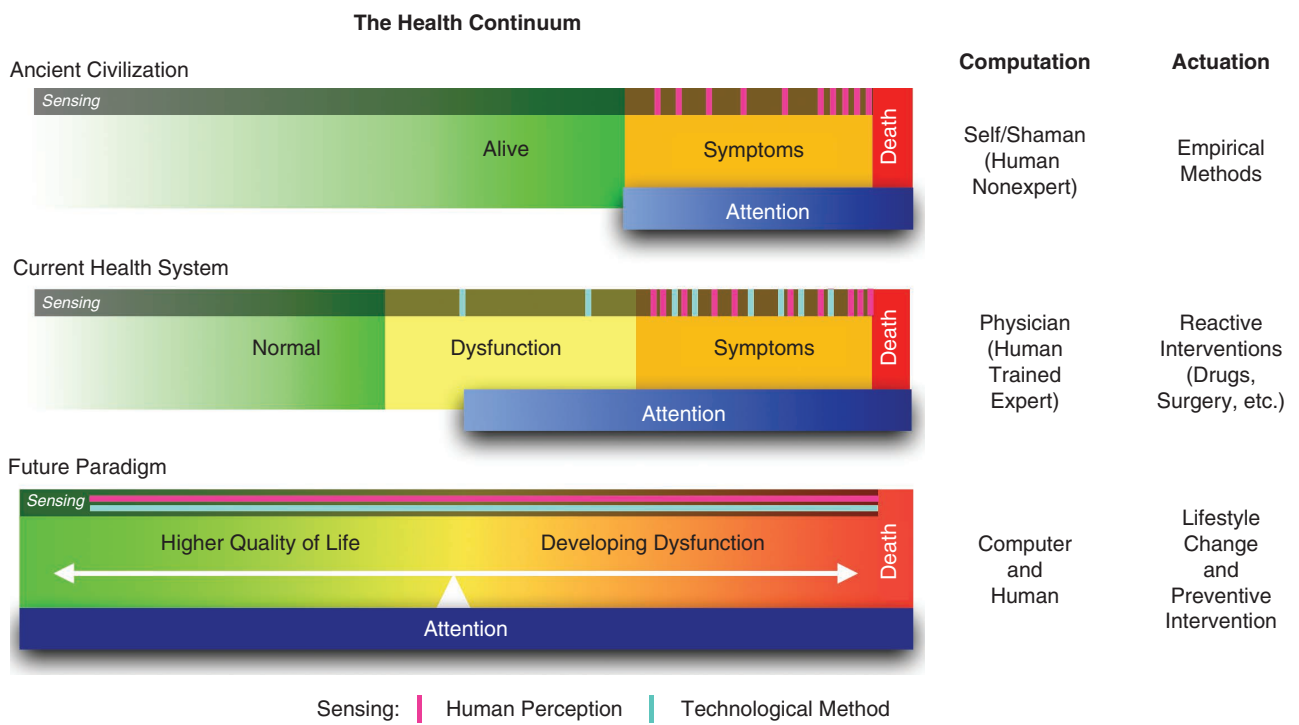

**FIGURE 1.** Once people know their health state is changing, they pay attention and take action. Hence, people take medication for pain (perceptual sensing) or high blood pressure (not easily perceived and requires modern sensor measurements). Understanding health states increases with the advent of sensors that can understand users' biology better than the individuals can feel it. Furthermore, the continuous–sensing approach escapes the traditional classification view of health and shifts to a continuous and quantitative view.

The purpose of personal health navigation (PHN) is to help an individual reach and maintain his or her desired health state (Figure 2). Systems that provide health navigation are broad, and humans may be integral components. Examples include a personal device such as a mobile phone and an entire Internet of Things (IoT) ecosystem within a building, a human trainer at a gym, or an advanced cancer center. PHN starts with a user-specified health goal. From there, the system starts to collect measurements about the person's health that relate to his or her goal to estimate current health states. To reach the desired goal optimally, the system breaks the steps from the current state and provides the next step to an actuation mechanism. All actions, whether advised by the system or not, are constantly measured to provide a new estimation of the patient's health state. A change in the health state updates the next action to be executed. A cycle of these actions moves the user's health state closer to the target health state. Upon reaching the destination, the system continuously ensures minimized deviation from that state. The vast unknown nature of health means that more diverse multimodal data will be obtained in the future and that this phenomenon requires combining these data sources.

### Human sensors

Two sets of perception influence our understanding of mental and physical health. Internal sensors capture feelings of perceived pain, energy, proprioception, and mood, to name a few. External sensors capture information about the outside world through sight, hearing, touch (including sensory pain, temperature, texture, and so on), taste, and smell. Humans use both sets to understand health states. Sensors measure a physical attribute to understand the physical world. Unfortunately, the ways our natural biological sensors unify, store, quantify, and use these data are not easily compatible with modern computing. Tedious methods that do not capture data continuously are used to share these two sets of sensory attributes well enough to optimize long-term health. Augmenting the external senses with tools such as imaging and microphones (for example, stethoscopes in the case of doctors) helps to feed data to computers, but the next data capture level is to include artificial sensors that go beyond our human senses.

### Artificial sensors

Throughout history, quantifying health at a higher resolution has always changed what it means to be considered healthy. The stethoscope, blood pressure cuff, and microscope all modified the definition. Continuous sensors, digital interactions, and biological measurements produce vast amounts of data. Smartphones collect steps, smart watches record heartbeats, and location tracking can obtain local pollution levels, while social media, shopping, and search history offer insights into personal interests and relationships.

Transitioning further into the information age, four key information sources are identified: perceptual, physical, biologic, and digital. Perceptual quantification takes human sensory information and provides continuous streams of data about emotions, sensory inputs, and brain activity. Physical sensors track the external environment to understand what a person's surroundings and behaviors are like. Biological sensors report on internal changes. Modern biological sensing mechanisms include "omic" data (genomic, microbiomic, and metabolomics), wearables to track physiology, the IoT and ambient devices, biomarkers, advanced imaging, and electrical signals. Digital sensors include interactions with all computers, phones, wearables, and other computing devices. This is an incomplete set of sensing mechanisms; we will continue to see an explosion of sensing technology in the 21st century.

### Sensor integration

To track the health continuum, measurements must be synchronized into a life-log database like Personicle.[4] Data streams are combined with event streams for semantic retrieval and understanding a person's daily life and environment.[5,6] Processing, analysis, compression, and safe yet useful storage with different semantics may also be very challenging. As modern sensors constantly stay connected to networks, they will have a continuous pulse on a rich variety of information about each person.

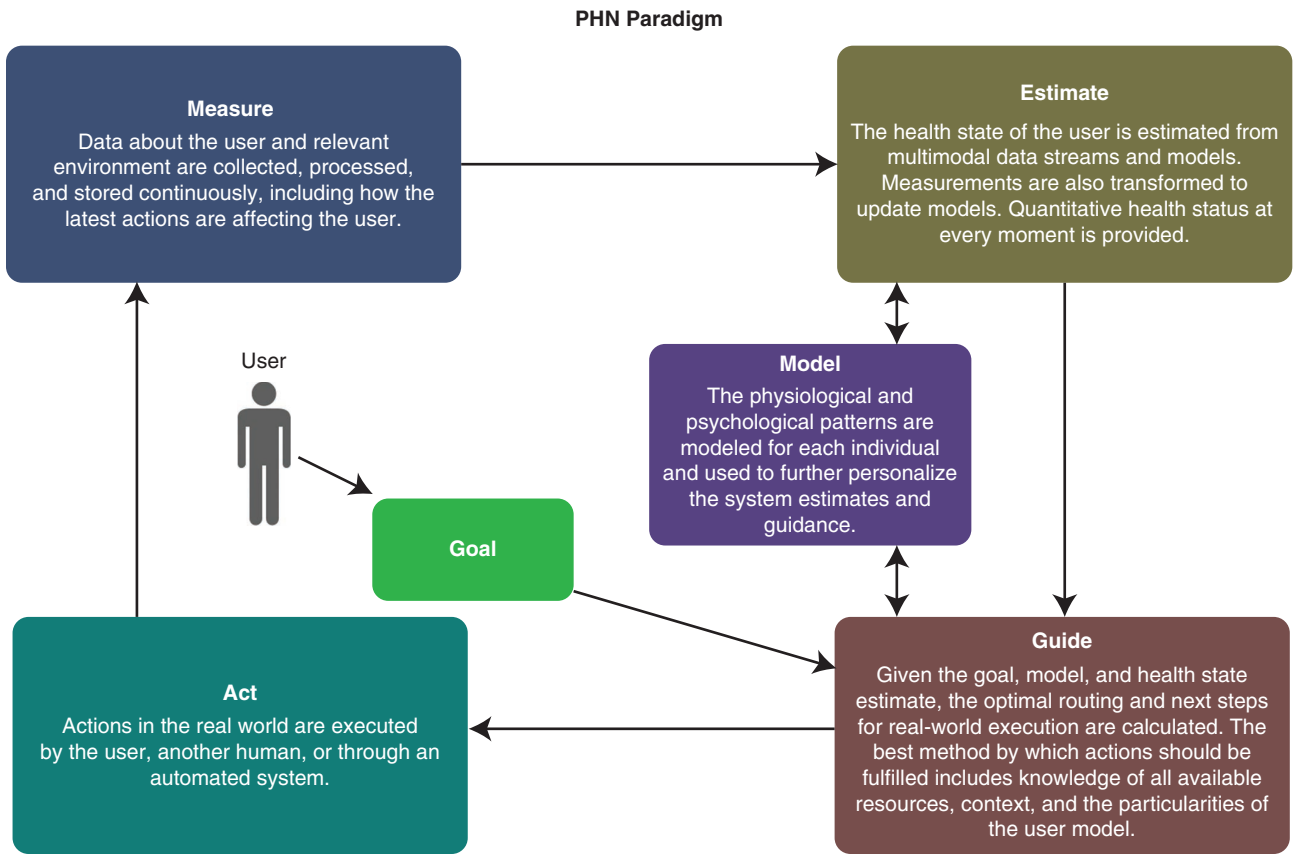

**FIGURE 2.** At a high level, PHN employs a cybernetic problem-solving approach to continuously move a user's health state toward a desired goal.

The cost of all sensing mechanisms is not equal, so data stream continuity and availability will vary. For example, a wearable heart-rate monitor is an inexpensive and continuous way to measure heart activity but will be less insightful than cardiac perfusion testing with a computerized tomography scanner to evaluate heart attack risk. The expensive sensor may be used to initialize or calibrate a health state, and the less expensive sensors, which are more continuous, would be available to estimate the current state from the most recently calibrated state.

### Estimate

Measurements are useful for estimating current health status by using learning in conjunction with reasoning (especially referencing domain knowledge) techniques. State estimations must take the entire evolving situation of an individual and his or her social connections into account, not just a static measurement, to understand state trajectories, resources, and problems. This estimation could indicate its proximity to biological dysfunction or its capacity for a desired goal. Two primary challenges arise in health estimation.

**Personal health state space.** Each person has a unique and finite possibility of biological states. Classic examples such as height, eye color, or fingerprints have a very small set of states after reaching adulthood. However, many aspects of biology have a large range of possible states, which change based on daily life. Consider how exercise changes the mitochondrial content of cells, how nutrition alters our metabolic processes, how medications modulate our blood pressure, and how ultraviolet light damages DNA. The changes in these states are what determines the movement on the health continuum. Mapping the personal health state space (PHSS) in totality is a complex, multidimensional challenge.

A pinnacle of health estimation research would be to aim to describe the PHSS in terms of a purely biological function, independent of pathology as the primary motivator. Layers on top of this coordinate system would contain all relevant health domain knowledge (Figure 3), similar to physical maps that use latitude, longitude, and altitude to describe the globe independently of any knowledge layers. Information layers such as roads, oceans, country borders, and satellite imagery allow for navigation within the space, depending on the context (driving requires roads and traffic layers, while flight requires air class and airport layers). Thus, geographic information systems have evolved to address these computational issues. Similarly, formal PHSS models and systems will emerge over time with the assimilation of knowledge that is associated with observability and controllability conditions.[2] A semantic translation to the relevant PHSS will also depend on the user's goals and system capabilities. For example, mapping the coordinates of a "healthy pregnancy" or "fast marathon run" in the PHSS requires matching relevant sections and coordinates that are of interest.

**Health state location.** There is constant flux within an individual's PHSS. To be successful, navigation must be able to assign an accurate location within the PHSS for any control decision, just as GPS provides physical world navigation. Understanding this location in real time, from multimodal data sources, will provide a highly relevant health estimation for a variety of applications. For example, monitoring a cardiovascular health state is useful to both endurance athletes and heart disease patients.[7] Estimation techniques have been of great interest for designing many applications, but health applications will require increasingly deep biological knowledge layers to define and estimate health states.

### Personal model

To predict the future, provide guidance, and understand the preferences and particularities of an individual, researchers must build personal models.[8] These models, which establish the premise that each individual is a unique system, are needed to best estimate the user's health state and how various inputs uniquely move them in the state space. Models are then accessed to provide precise guidance. The personal model can build knowledge and predict many aspects of an individual's life, such as how a person reacts to different stimuli under specific conditions or physiological change from an intervention.

To model a person, a combination of long- and short-term information must interact. Long-term models of an individual can come from the genome or the history of event patterns and behavior. Biological or high-resolution continuous sensors can capture short-term models. These models are not static; they continuously change with an individual's age, life events, and other life parameters, which makes model-building a dynamic and causal understanding process.[9] Models must consider, at least, the following.

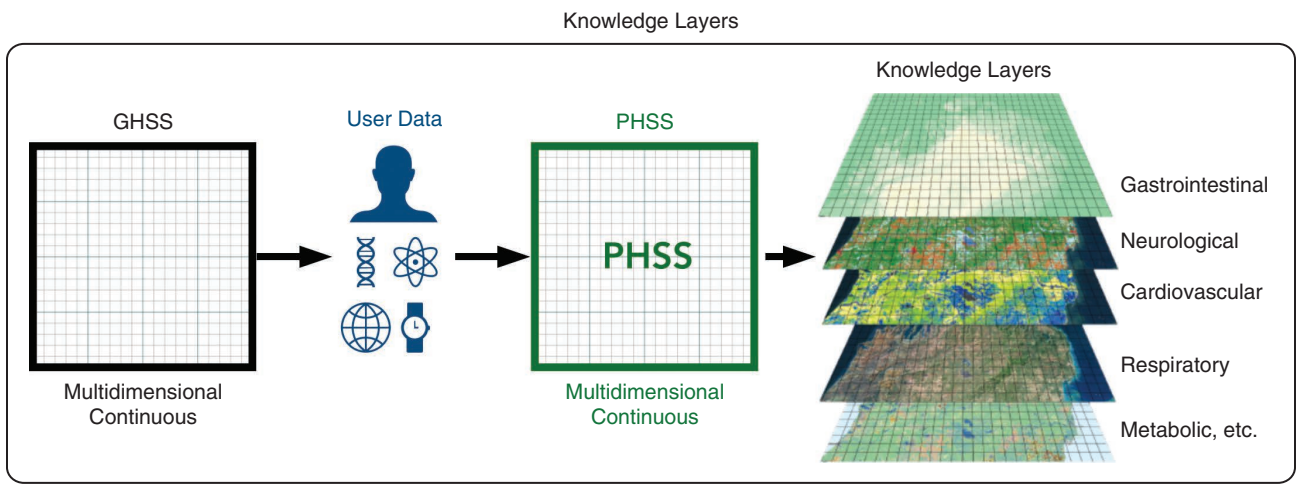

**FIGURE 3.** The knowledge layers in PHN require mapping the semantic information from health and biology to the PHSS, which allows the translation of semantics, understanding of the health state, and routing for navigation.

**Physiologic function.** Dynamic health attributes include biological health, such as organ function and capacity (cardiorespiratory capacity, bone strength, and metabolism). Models may have various time and size levels in terms of detail, depending on the application. Short-term models can already predict how electrical activity in the heart may produce an arrhythmia,[10] while a long-term model may predict the impact of long-term lifestyle habits on developing atherosclerosis. Cellular or gross organ-level processes can be modeled separately or in combination. These functions constantly change based on daily factors that include sleep, nutrition, medication, or stress.

**Mental condition.** Extrapolating and modeling mental health will be a complex challenge that requires advanced sensing mechanisms such as affective computing. Mental health components could include various time resolutions of interest, such as overall mental stability or minute-by-minute mood analysis. Evaluating risks for erroneous modeling prior to field deployment will be essential for the users' safety.

**Behavior.** Understanding user preferences, influences, and motivations will be essential to close the loop when humans are involved as actuators. The system uses this information to tailor guidance in a way that makes living a healthy lifestyle pleasurable and convenient, enhancing the quality of life. To be effective, irrational behavior, social tendencies, and situations in which temptation overrides true long-term goals must be included in the model.[11] In addition, risky behaviors may warrant help to prevent harmful events such as drunk driving.

**Risk.** Researchers can model risk assessment for reversible and irreversible physiologic dysfunction by using basic biological models. An example of disease risk would be the chance that an individual would develop metabolic syndrome, a condition that medical professionals could possibly reverse with an early-stage intervention. An example of a damage risk would be the likelihood of this syndrome turning into type 1 diabetes, resulting in permanent pancreatic damage. Risk modeling will also include behavior and mental condition models to understand the propensity for self-harm, accidents, or harm to others, as mentioned previously. To provide the greatest value, resources are prioritized and invested based on evaluating the likelihood and consequence of these combined risks.

## Guide

After passively monitoring, estimating, and modeling an individual, the next step in navigation would be to input a goal and receive guidance on the next step needed to reach that goal. In the case of health, a user selects a desired semantic state, and then the system begins to navigate by decomposing the goal to intermediate states, routing to the next state, and deciding the best execution strategy for actuation.

**Goal decomposition.** To reach the next intermediate state, goal decomposition will require long-term planning of the user's goal state within PHN and translating that goal state into short-term goals for guidance. At first, the system will need to translate the semantic goal to the PHSS. Because the PHSS is complex, a given macro health goal may need to be broken into several intermediate states and subgoals that the system can focus on reasonably. There may be a set of short-term goals at any given moment, each with its own set of intermediate states and subgoals. At the lowest atomic level for an application, PHN lists fulfillment tasks in priority order so that users can execute them in the real world to attain a subgoal or the next intermediate state.

Using the example goal of a "healthy pregnancy," experts would divide it into six parts: preconception, first trimester, second trimester, third trimester, delivery, and postpartum. Each of these intermediate steps will have various intermediate states and subgoals that are required for optimization. Focusing on preconception as an example, the mother must have adequate nutrition and fertile ovulation. Each subgoal will be defined by a smaller state space. In the case of nutrition, models will list macro (carbohydrates, protein, fat) and micro (vitamins, minerals) nutrients. These components may then be the chosen atomic-level intermediate states for applying nutrition.

**Routing.** Making a route on a map requires not only knowing the start and end points but also all the layers of roads and traffic. In the case of PHN, each set of intermediate states and subgoals will have its own layer of information, which is relevant for mapping, along with costs and constraints to transition among intermediate states. Interactions in PHN will be extremely complex due to its large dimensionality. Contradictions between various user goals must be handled also, perhaps through prioritization or weighting. Means–ends analysis or other problem-solving techniques, along with appropriate routing algorithms, will reveal the best intermediate states for the user to reach the goal.

In the case of food, it is necessary to map available food options, nutrition, and locations, along with how each choice can fulfill the tasks demanded. Consider models for a breakfast, lunch, snack, and dinner in the daily subgoal of a nutritional state. The total components must be coordinated to reach the subgoal appropriately. For example, if breakfast does not provide enough iron, then it must be compensated for in other meals. Researchers must further consider the logistics of adding a dish into the equation. Hence, solving routing for various PHN components remains an open opportunity.

**Execution strategy.** Context-aware recommendations will be critical for understanding how to provide navigational guidance.[12] In daily life, transitory situations present constant opportunities to execute the next step in fulfilling a subgoal and thus to help reach the next intermediate state. For a particular user, identifying these contexts through continuous sensing is the best execution strategy.

Referring to the nutritional case, the system will need to decide the best meal choice to move the user's health state closer to his or her goal state based on situational constraints. This decision will require knowledge of all available meals, as well as their locations, hours, nutrition facts, and ingredients along with the personal sensor data modeling the physiologic needs.

**Act**

Computing optimal guidance is not enough. To have a meaningful impact, guidance must be situationally actionable and implemented in the real world. A spectrum of actuation ranging from fully automated systems to human-fulfilled actions is given for the system to be effective in the physical world.

**Machine driven.** Many machines automate tasks such as laundry or aircraft control, which are instances of completely automated control mechanisms that can make an action in the real world control a task. For type 1 diabetes, continuous glucose monitoring paired with insulin pumps can control blood sugar far better than humans can. In the future, actuators of all types will be pervasive for health-related actions in everyday life. Modulating the environment is a prime example. Imagine homes with automated biological circadian rhythm control via connected lights, digital screen color variation, thermostats that mimic natural temperature fluctuations, and motorized curtains with smart white noise. In these instances, the executed actions are completely automated, and no human intervention or uncertainty is involved, so these cases can be considered deterministic actuators in PHN.

**Human driven.** Humans are not always good at understanding complex instructions or remembering them at the right time and in the right situations. They also are easily influenced and distracted, which makes them nondeterministic actuators. Even when the user fully intends to reach a health goal, adhering to healthy routines can be difficult.

The primary challenging task is to understand how to influence people at the right time and with the best medium to produce a desired action. For each individual, actions must be persuasive in the given context, practically feasible, and encourage participation. Richard Thaler won the 2017 Nobel Prize for understanding the psychological underpinnings and control of the decision-making process in humans.[11] Specifically, the following three high-impact contributions interweave directly with health navigation.

1. *Limited rationality*: Simplified decision-making focuses on the narrow impact of each individual decision rather than on its overall effect. Computing these complex effects can offload these tasks from the user.
2. *Lack of self-control*: A planner–doer model describes the internal tension between long-term planning and short-term doing. Succumbing to short-term temptation is an important reason that plans to make healthier lifestyle choices often fail. By quantifying the health cost or benefit in any situation, this tension can be reduced when users can make more informed decisions.
3. *Nudging*: By using subtle cues, users are pushed to make a certain decision without force, thus

avoiding the pitfalls of self-control and limited rationality. Systems that employ these tactics can unobtrusively steer choices, leading to the goal state.

As PHN is able to better model the behavior of an individual, the probability of executing a given action can be increased using the considerations listed. These improvements require close collaboration with psychologists and social scientists. To solicit trust and further participation, the system explains why it suggests the nudge or action to users, and it offers alternate options to embolden the feeling of choice. Another avenue for human actuation will be to reduce the decision burden by eliminating the tediousness involved in being healthy. Even professional athletes do not measure every gram or determine every ingredient in their food due to the information and calculation burden. Translated into simple, step-by-step, context-aware guidance, users can better fulfill PHN actuation steps. PHN can start to apply control to the health state, either with fulfillment by a machine or with user participation, to move the health state toward the goal and close the loop.

## THE ROAD AHEAD: SECURITY, PRIVACY, AND ETHICAL CHALLENGES

Research efforts have focused on developing various components of PHN (Figure 4). In this section, we outline some key challenges for research and implementation.

PHN systems must stay focused on assisting users. Ethical standards assign individuals freedom of choice in health matters. Users must also control and consent to share their data after understanding how that information will be used, especially in relation to their own future and relationships with family, employers, and the government. Ethical protections should be put in place against self-incrimination and discrimination. Conversely, users who are willing to anonymously share data for scientific progress should be able to do so easily. Given the sensitive and identifying nature of some data streams such as location, genetics, and behaviors, sharing data and insights

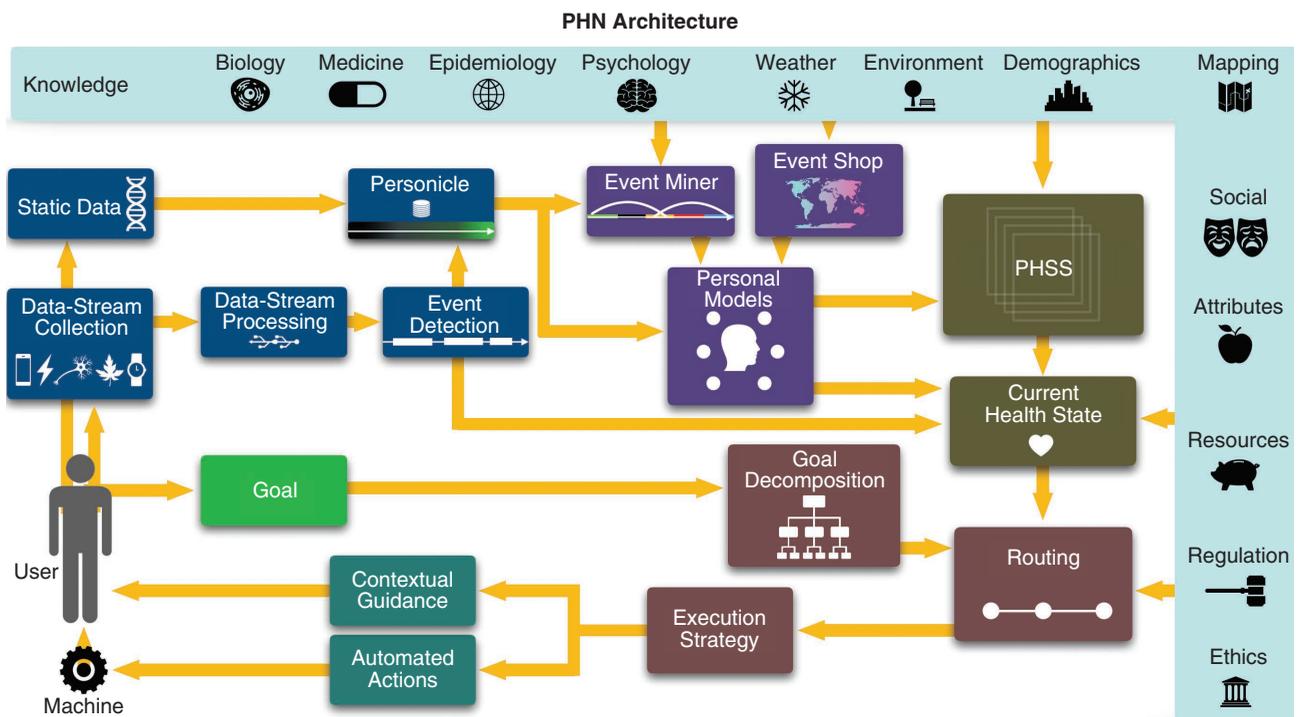

**FIGURE 4.** The PHN architecture reflects all the major computing elements described in this article working together to continuously provide navigation. Many modules have been the subject of active research in the past decades.[4, 6–10, 12, 13]

will require unique approaches to uphold these ethical requirements.

Policy must follow these requirements to retain user trust. Accountability for data creation, usage, and rights for various stakeholders need to be further explored. In addition, PHN developers will be directly affecting the health of users, putting them in a parallel position with responsibilities similar to a physician or medical device. Transparency in how and why the system provides navigation will require research into AI explainability.[9] Data bias, application range, and error potential must be understood before being implemented with users. Enforcement mechanisms to ensure trust should use emerging methods such as distributed ledger systems, biometric authentication, advanced encryption, and protection against data reverse inference.

Future systems can leverage modern computing power with rich multimodal data to understand movements on the health state space continuously and to contextually guide users toward wellness. This nexus of biomedical knowledge, sensor technology, computing power, mobile networks, and artificial intelligence will fuel major growth in PHN systems.[14] The European Union and the United States approved the first PHN mobile application for contraception in 2018.[15] This fluid framework can advance the promise of predictive, personalized, and precise health care in the future.

## ABOUT THE AUTHORS

**NITISH NAG** is an M.D./Ph.D. candidate at the University of California, Irvine, in the Medical Scientist Training Program, where he is pursuing an M.D. in conjunction with a Ph.D. in computer science. His research interests include developing future technologies in the areas of health wellness and preventive medicine. Nag received a triple major (summa cum laude) from the University of California, Berkeley, in biochemistry, nutrition metabolism, and integrative biology. Contact him at nagn@uci.edu.

**RAMESH JAIN** is a Donald Bren professor in information and computer sciences at the University of California, Irvine, where he is currently researching future health computing. He is the recipient of several awards including the ACM SIGMM Technical Achievement Award 2010. He is a Fellow of the IEEE, Association for Computing Machinery, Association for the Advancement of Artificial Intelligence, International Association for Pattern Recognition, and International Society for Optics and Photonics. Contact him at jain@ics.uci.edu.